\documentstyle{article}
\input epsf

\newfont{\SETT}{msbm10 scaled \magstep4}
\newfont{\SET}{msbm10 scaled \magstep3}
\newfont{\Set}{msbm10 scaled \magstep2}
\newfont{\Settoc}{msbm10 scaled \magstep1}
\newfont{\set}{msbm10}

\newcommand{\beq}{\begin{equation}}
\newcommand{\eeq}{\end{equation}}
\newcommand{\bea}{\begin{eqnarray}}
\newcommand{\eea}{\end{eqnarray}}

\newcommand{\inx}{\int d^{2}{\rm{\bf x}}\, }

\newcommand{\x}{{\rm{\bf x}}}
\newcommand{\y}{{\rm{\bf y}}}
\newcommand{\z}{{\rm{\bf z}}}
\newcommand{\rv}{{\rm{\bf r}}}
\newcommand{\jv}{{\rm{\bf j}}}
\newcommand{\A}{{\rm{\bf A}}}

\newcommand{\k}{\widehat{{\rm{\bf k}}}}
\newcommand{\zv}{{\rm{\bf 0}}}

\newcommand{\cd}{\partial}
\newcommand{\D}{\nabla}
\newcommand{\CD}{{{\bf D}}}

\newcommand{\dxy}{\delta(\x-\y)}

\newcommand{\kn}{K_{0}}
\newcommand{\ko}{K_{1}}
\newcommand{\R}{\mbox{\set R}}

\newcommand{\C}{\mbox{\set C}}

\newcommand{\hf}{\frac{1}{2}}
\newcommand{\ol}{\overline}

\newcommand{\thet}{\widehat{\mbox{\boldmath $\theta$}}}

\newcommand{\th}{\vartheta}

\newcommand{\square}{\Box}

\begin{document}
\title{Static Intervortex Forces}
\author{J.M. Speight \\
Department of Mathematics \\
University of Texas at Austin \\
Austin, Texas 78712, U.S.A.}
\date{}
\maketitle

\begin{abstract}

A point particle approximation to the classical dynamics of well separated 
vortices of the abelian Higgs model is developed. A static vortex is
asymptotically identical to a solution of the linearized field theory
(a Klein-Gordon/Proca theory) in the presence of a singular point source
at the vortex centre. It is shown that this source is a composite scalar
monopole and magnetic dipole, and the respective charges are determined
numerically for various values of the coupling constant. The interaction
potential of two well separated vortices is computed by calculating the 
interaction Lagrangian of two such point sources in the linear theory.
The potential is used to model type~II vortex scattering.

\end{abstract}

\section{Introduction}
\label{sec:int}

The abelian Higgs model \cite{Hig} is a relativistic field theory consisting
of a complex scalar field $\phi$ coupled to a $U(1)$ gauge
field $A_{\mu}$, and given a Higgs symmetry-breaking self
interaction which allows topologically stable solitons to
exist. The present work concerns the $(2+1)$-dimensional model, in which the
solitons are simple lumps of energy called vortices (in $(3+1)$ dimensions
the model admits extended string-like solitons \cite{Vil}
 whose gravitational effects may
be important
in early Cosmology). Multivortex dynamics falls into one of three
regimes, depending on the Higgs mass $\mu$: if $\mu$ is small, vortices attract
one another (the type I regime), while if $\mu$ is large they repel (type II),
and at one critical value of $\mu$ static vortices exert no net force on one
another. 

In this paper, we calculate the interaction potential of two widely
separated vortices by means of a novel approximation. The 
idea is that,
viewed from afar, a static vortex looks like a solution of a linear field
theory in the presence of a singular point source at the vortex centre. As
will be shown, the appropriate point source is a composite scalar monopole
and magnetic dipole in a Klein-Gordon/Proca theory. If physics is to be model 
independent,
then the forces between well separated vortices should approach those between
the corresponding point particles in the linear theory as the separation grows.
Proceeding on this assumption, we calculate the asymptotic static two vortex
potential. Our answer agrees with that of Bettencourt and Rivers, which was
derived using a field superposition ansatz \cite{BetRiv}.

The monopole charge and dipole moment of the composite point source depend on
the Higgs mass $\mu$. To fix their values, one must solve the static nonlinear
field equations with vortex boundary conditions, and such solution is
perforce numerical. Using a simple Runge-Kutta scheme we have determined
these charges for various values of $\mu^{2}$. Here our results disagree with
the work of Bettencourt and Rivers, who, while leaving undetermined the
constants analogous to these charges, make certain assumptions about them
which appear ill justified.

The asymptotic potential reproduces the aforementioned dynamical trichotomy
into type~I, type~II and critical regimes found in the nonlinear model. As
an application, the scattering of type~II vortices is calculated and compared
with numerical simulations.

\section{The abelian Higgs model}
\label{sec:ahm}

We begin by reviewing some standard results concerning the abelian Higgs model
\cite{Hig}.
 The Lagrangian density is
\beq
\label{1}
{\cal L}=\hf D_{\mu}\phi\ol{D^{\mu}\phi}
-\frac{1}{4}F_{\mu\nu}F^{\mu\nu}-\frac{\mu^{2}}{8}(|\phi|^{2}-1)^{2}
\eeq
where $D_{\mu}\phi=(\cd_{\mu}+iA_{\mu})\phi$ is the gauge covariant derivative,
$F_{\mu\nu}=\cd_{\mu}A_{\nu}-\cd_{\nu}A_{\mu}$ is the field strength tensor, 
and $\R^{2+1}$ has signature $(+,-,-)$. Note that the electric charge and 
vacuum magnitude
of the Higgs field have been normalized to unity, leaving only one parameter
$\mu,$ the Higgs mass. With these conventions the model is critically
coupled if $\mu=1$, and in the type~I (type~II) regime if $\mu<1$ ($\mu>1$). 
The Euler-Lagrange equations derived from ${\cal L}$ are 
\bea
D_{\mu}D^{\mu}\phi-\frac{1}{2}\phi(|\phi|^{2}-1)&=&0 \nonumber \\
\label{1.5}
\cd_{\mu}F^{\mu\nu}+\frac{i}{2}(\phi\cd^{\nu}\bar{\phi}
-\bar{\phi}\cd^{\nu}\phi)
+|\phi|^{2}A^{\nu}&=&0,
\eea
a set of coupled, 
nonlinear, hyperbolic partial differential equations of 
which no nontrivial solutions are known. 

If a
configuration is to have finite energy, the fields should satisfy the following
boundary conditions as $r=|\x|\rightarrow\infty$,
\beq
\label{9}
|\phi| \rightarrow 1,\qquad \CD\phi \rightarrow \zv,
\eeq
whence $\phi_{\infty}:=\lim_{r\rightarrow\infty}\phi$
takes values on the unit circle in $\C$ and is thus a continuous map 
$S^{1}_{\infty}\rightarrow S^{1}$, where $S^{1}_{\infty}$ represents the
circle at spatial infinity. Such configurations fall into disjoint homotopy
classes labelled by the degree of $\phi_{\infty}$ (an integer, $n$, also
called the winding number of $\phi$). The time evolution defined by the field
equations conserves energy, and so cannot take a field in one homotopy class 
into a different class, since such an evolution, being continuous, would
define a homotopy between distinct classes, a contradiction. Hence $n$ is a
topologically conserved quantity.

Another consequence of the boundary conditions is that the total magnetic
flux is topologically quantized, for
\beq
-\inx F_{12}=2n\pi.
\eeq
follows from (\ref{9}) and Stokes' theorem.
A static solution with $n=1$ is a stable lump of energy called a vortex.
This may be visualized as a single flux tube in $\R^{3+1}$ with translation
symmetry along the $x^{3}$ axis. It has a total flux of $2\pi$ penetrating the
physical plane.

\section{Vortex asymptotics}
\label{sec:va}

The first task in the point particle approximation is to find out what a static
vortex looks like far from its core \cite{BetRiv,NieOle}. We place the vortex 
at the origin, and use
plane polar coordinates. Substituting the ansatz
\bea
\phi &=& \sigma(r)e^{i\theta} \nonumber \\
\label{12}
(A_{0},A_{r},A_{\theta}) &=& (0,0,-a(r)) 
\eea
($\sigma$ is real) the field equations (\ref{1.5}) reduce to two coupled 
nonlinear
ordinary differential equations,
\bea
\frac{d^{2}\sigma}{dr^{2}}+\frac{1}{r}\frac{d\sigma}{dr}
-\frac{1}{r^{2}}\sigma(1-a)-\frac{1}{2}\mu^{2}\sigma(\sigma^{2}-1)&=& 0
\nonumber \\
\label{14}
\frac{d^{2}a}{dr^{2}}-\frac{1}{r}\frac{da}{dr}
+(1-a)\sigma^{2}&=& 0,
\eea
the equations for $A_{0}$ and $A_{r}$ being trivially satisfied. Regularity
demands that $\sigma(0)=a(0)=0$ while the boundary conditions 
(\ref{9}) become
\beq
\label{15}
\lim_{r\rightarrow\infty}\sigma(r)=\lim_{r\rightarrow\infty}a(r)=1.
\eeq
Note that the ansatz has unit winding by construction. No exact solutions
of (\ref{14}) with these boundary conditions are known, but numerical solutions
suggest that both $\sigma$ are $a$ are monotonic functions of $r$, and that
the ansatz produces an isolated lump like structure.

We are interested in the asymptotic forms of $\sigma$ and $a$, and for these
explicit expressions do exist. Define the functions $\alpha$ and $\beta$ such
that
\beq
\label{16}
\sigma(r)=1+\alpha(r),\quad a(r)=1+\beta(r).
\eeq
Then (\ref{15}) implies that $\alpha$ and $\beta$ are small at large $r$,
so we substitute (\ref{16}) into (\ref{14}) and linearize in $\alpha$ and
$\beta$,
\bea
\mu^{2}\left(\frac{d^{2}\alpha}{d(\mu r)^{2}}
+\frac{1}{\mu r}\frac{d\alpha}{d(\mu r)}-\alpha\right)&=&0 \nonumber \\
\label{17}
\frac{d^{2}\,\,}{dr^{2}}\left(\frac{\beta}{r}\right)
+\frac{1}{r}\frac{d\,\,}{dr}\left(\frac{\beta}{r}\right)
-\left(1+\frac{1}{r^{2}}\right)\frac{\beta}{r} &=& 0.
\eea
These are the modified Bessel's equations of zeroth order for $\alpha$ in
$\mu r$ and first order for $\beta/r$ in $r$ respectively. Hence, at large
$r$,
\bea
\alpha &\sim& \frac{q}{2\pi}K_{0}(\mu r) \nonumber \\
\label{18}
\beta &\sim& \frac{m}{2\pi}rK_{1}(r),
\eea
where $K_{n}$ is the $n$--th modified Bessel's function of the second kind
\cite{AbrSte}. Note that $K_{1}\equiv-K_{0}'$.

Since we have linearized the field equations, the asymptotic solutions contain
unknown scale constants $q$ and $m$ which can only be fixed by solving
(\ref{14}) numerically. Rather than solving the boundary value problem
$\sigma(0)=a(0)=0$, $\sigma(\infty)=a(\infty)=1$, we solve the initial value
problem $\sigma(0)=a(0)=0$ using $\sigma'(0)$ and $a'(0)$ as shooting 
parameters. In fact, this is a slight oversimplification: due to the 
singularities of equations (\ref{14}) at the origin, we must shoot from 
$r=r_{0}$, some small positive number, rather than $r=0$. Substituting Taylor
expansions for $\sigma$ and $a$ into (\ref{14}) we find that, near the
origin,
\bea
\sigma&=&a_{1}r+\frac{1}{4}a_{1}\left(b_{2}+\frac{\mu^{2}}{4}\right)r^{3}
+O(r^{5}) \nonumber \\
\label{19}
a&=&b_{2}r^{2}-\frac{a_{1}^{2}}{8}r^{4}+O(r^{5}).
\eea
We use $a_{1}$ and $b_{2}$ as shooting parameters, adjusting them until the
numerical solution has $\sigma(r_{\infty})\approx 1\approx a(r_{\infty})$,
where $r_{\infty}$ is some large positive number, the effective infinity.
Having generated such a numerical solution, we compare it at large $r$ to the
asymptotic forms (\ref{18}) and deduce $q$ and $m$.

The results of this procedure using a fourth order Runge-Kutta method with
$r_{0}=10^{-8}$ and $r_{\infty}=10$ for various values of $\mu^{2}$ are
presented in table 1. That $r_{\infty}$ is so small is unfortunate but 
necessary: at large $r$ the field equations reduce to Bessel's equations,
which have two independent solutions, one exponentially decaying and the
other exponentially growing. We seek to pick out the former and completely
exclude the latter, an impossible task. Hence, all numerical solutions blow
up at large $r$, and even though $a_{1}$ and $b_{2}$ were tuned to six
decimal places, the Runge-Kutta algorithm could not shoot beyond $r=10$. 

\begin{table}
\begin{center}
\begin{tabular}{|c|c|c|c|} \hline
$\mu^{2}$ & $q$ & $m$ & $r_{c}$  \\ \hline
0.4 & -7.54 & -14.92 & 4.23   \\
0.6 & -8.71 & -12.61 & 3.75  \\
0.8 & -9.70 & -11.31 & 3.43 \\
0.9 & -10.14 & -10.89 & 3.22 \\
1.0 & -10.58 & -10.57 & -    \\
1.1 & -10.98 & -10.31 & 2.98 \\
1.2 & -11.43 & -10.06 & 3.07 \\
1.3 & -11.80 & -9.85 & 2.96  \\
1.4 & -12.23 & -9.66 & 2.95  \\
1.6 & -13.04 & -9.34 & 2.88  \\
1.8 & -13.97 & -9.09 & 2.87  \\
2.0 & -14.50 & -8.86 & 2.72  \\ \hline 
\end{tabular}
\end{center}
\caption{{\sf Numerical values of vortex scalar charge $q$ and magnetic dipole 
moment $m$. The other data are the critical points  
of the static intervortex potential.}}
\label{table}
\end{table}

Nevertheless, the qualitative nature of the $\mu^{2}$ dependence of $q$ and
$m$ is clear. In particular, two points about the numerical charges are 
noteworthy. First, at critical 
coupling ($\mu^{2}=1$), $q\approx m$. In fact, one can prove that $q\equiv m$
exactly in this case, because the static $\mu^{2}=1$ vortex satisfies
a pair of {\em first} order field equations. These are deduced by means of an
argument due to Bogomol'nyi \cite{Bog}, 
\bea
(D_{1}+iD_{2})\phi &=& 0  \\
\label{BOG}
-F_{12}+\frac{1}{2}(|\phi|^{2}-1)&=& 0
\eea
(in the $A_{0}=0$ gauge) and within our ansatz take the form
\cite{Sam},
\bea
\label{bog}
r\frac{d\sigma}{dr}-(1-a)\sigma &=& 0  \\
\frac{2}{r}\frac{da}{dr}+(\sigma^{2}-1)&=&0. 
\eea
Substituting (\ref{16}) into (\ref{bog}) yields, on linearizing,
\beq
\beta=-r\frac{d\alpha}{dr}
\eeq
and so $\alpha=qK_{0}(r)/2\pi\Rightarrow\beta=qrK_{1}(r)/2\pi$. Thus 
$m\equiv q$. We emphasize that this argument works {\em only} at critical
coupling. 

Second, $|q|$ and $|m|$ are monotonic functions of 
$\mu^{2}$, $|q|$
increasing and $|m|$ decreasing. Bettencourt and Rivers \cite{BetRiv} also
find the asymptotic forms (\ref{18}), but leave their charges analogous to
$q$ and $m$ undetermined. For purposes of calculation, they make two
assumptions about the charges which, in the light of table 1, may prove
ill-justified. First, they assume that $q=m$ is approximately true away from
$\mu^{2}=1$, whereas in our results, $q/m$ varies between $0.50$ and $1.64$.
Second, they impose the condition that the magnetic flux of a vortex should
vanish at $r=0$ and deduce that $m=-2\pi$ (it is unclear why a condition at
$r=0$ should directly constrain the asymptotic behaviour as $r\rightarrow
\infty$). This result may be valid for
very large $\mu^{2}$, but is certainly flawed close to $\mu^{2}=1$ since the
$\mu^{2}=1$ vortex has {\em maximum} magnetic flux at the origin, as is easily
seen from the lower Bogomol'nyi equation (\ref{BOG}). So, they combine 
assumptions which are
individually true only in widely disparate physical regimes.

\section{The point vortex}
\label{sec:pv} 

The next task is to replicate the vortex asymptotics, found above, in the
linear field theory by coupling the fields in standard fashion to a scalar 
density $\rho$ and a vector current $j_{\mu}$, as yet undetermined. To 
linearize the abelian Higgs model, we choose
gauge so that $\phi$ is real. Defining the field $\psi=1-\phi$, the vacuum is
then $\psi=0$, and the linear Lagrangian density is obtained by expanding 
(\ref{1}) up to quadratic order in $\psi$ and $A_{\mu}$,
\beq
\label{20}
{\cal L}_{\rm free}=\hf\cd_{\mu}\psi\cd^{\mu}\psi-\hf\mu^{2}\psi^{2}
-\frac{1}{4}F_{\mu\nu}F^{\mu\nu}+\hf A_{\mu}A^{\mu}.
\eeq
Including the external source Lagrangian density,
\beq
\label{21}
{\cal L}_{\rm source}=\rho\psi-j_{\mu}A^{\mu},
\eeq
we obtain the 
following massive, inhomogeneous wave equations for $\psi$ and $A_{\mu}$,
\bea
\label{22}
(\square+\mu^{2})\psi &=& \rho \\
\label{23}
(\square+1)A_{\mu} &=& j_{\mu}+\cd_{\mu}(\cd_{\nu}j^{\nu}).
\eea
All gauge freedom has been exhausted, and there is no global $U(1)$ symmetry
of ${\cal L}_{\rm free}$ with whose Noether current we can identify
$j_{\mu}$ because $\psi$ is real. Hence there is no reason to assume that
$j_{\mu}$ is a conserved current, and we cannot set the extra ``fictitious 
current'' term in the Proca equation (\ref{23}) to zero.

To make comparison with the asymptotic vortex fields, these must first be 
converted to the real $\phi$ gauge. Since $\phi$ has non-zero winding, there
is no gauge transformation regular on all $\R^{2}$ which will accomplish this.
However, we only require comparison at large $r$, so for our purposes it is
sufficient that the transformation be regular on $\R^{2}\backslash\{\zv\}$.
Since a singular point source will be introduced into the linear theory, this
is from the outset regular only on $\R^{2}\backslash\{\zv\}$. So, we unwind
the static vortex (\ref{12}) with gauge transformation 
$\phi\mapsto e^{-i\theta}\phi$, $A^{\mu}\mapsto A^{\mu}+\cd^{\mu}\theta$ to 
obtain
\bea
\label{24}
\phi&=&\sigma(r)\sim 1+\frac{q}{2\pi}K_{0}(\mu r) \\
\label{25}
A_{\theta}&=&-a(r)+1\sim\frac{m}{2\pi}rK_{1}(r),
\eea
while $A_{r}=A_{0}=0$. It is convenient to introduce a unit vector $\k$ in a
fictitious third direction perpendicular to the physical plane, so that
the $\R^{3}$ vector product can be defined. In terms of the 2-vector field
$\A$, the unwound asymptotic behaviour is
\beq
\label{26}
\A\sim-\frac{m}{2\pi}K_{0}'(r)\thet=-\frac{m}{2\pi}\k\times\D K_{0}(r).
\eeq
We thus seek sources $\rho$ and $j_{\mu}$ such that the solutions of
(\ref{22},\ref{23}) are
\bea
\label{27}
\psi&=&\frac{q}{2\pi}K_{0}(\mu r) \\
\label{28}
(A^{0},\A)&=&\left(0,-\frac{m}{2\pi}\k\times\D\kn(r)\right).
\eea

The static Klein-Gordon equation in $(2+1)$ dimensions has Green's function
$\kn$,
\beq
\label{29}
(-\Delta+\mu^{2})\kn(\mu r)=2\pi\delta(\x).
\eeq
Substituting (\ref{27}) into (\ref{22}) and using (\ref{29}) one finds that
\beq
\label{30}
\rho=q\delta(\x).
\eeq
Similarly, substitution of (\ref{28}) into (\ref{23}) yields
\beq
\label{31}
\jv-\D(\D\cdot\jv)=-m\k\times\D\delta(\x).
\eeq
Taking the divergence of (\ref{31}) one sees that $\D\cdot\jv$ is a solution of
the {\em homogeneous} static Klein-Gordon equation, so if $\jv$ is a point
source (meaning $\jv=\zv$ except at $\x=\zv$) then $\D\cdot\jv=0$ everywhere.
Thus the unique point source satisfying (\ref{31}) is
\beq
\label{32}
\jv=-m\k\times\D\delta(\x).
\eeq
Since $A^{0}=0$ we take $j^{0}=0$.
The physical interpretation of these expressions for $\rho$ and $\jv$ is that
the point source consists of a scalar monopole of charge $q$ and a magnetic
dipole of moment $m$ perpendicular to the physical plane. Both $q$ and $m$ are
negative (see table 1). We refer to this composite point source as the point 
vortex. 

\section{The static intervortex potential}
\label{sec:sip}

Having found the scalar charge and magnetic dipole moment carried by a point
vortex, it is straightforward to calculate the force between two such vortices
held at rest, in the framework of the linear theory. The interaction Lagrangian
for two arbitrary (possibly time dependent) sources $(\rho_{1},j_{(1)})$ and
$(\rho_{2},j_{(2)})$ is
\beq
\label{33}
L_{\rm int}=L_{\psi}+L_{A}=\inx \rho_{1}\psi_{2}
-\inx j^{\mu}_{(1)}A_{\mu}^{(2)}
\eeq
where $(\psi_{i},A_{(i)})$ are the fields induced by source 
$(\rho_{i},j_{(i)})$ according to the wave equations (\ref{22},\ref{23}). This
is found by extracting the cross terms in 
$\inx({\cal L}_{\rm free}+{\cal L}_{\rm source})$ where $(\rho,j)$ is the
superposition of the two sources, and $(\psi,A)$ is a superposition of the 
induced fields. The expression (\ref{33}) looks asymmetric
under interchange of sources $1\leftrightarrow 2$, but in fact $L_{\rm int}$
is symmetric as may be shown using the wave equations (\ref{22},\ref{23}) and
integration by parts.

Now consider the case of two static point vortices, vortex 1 at $\y$ and
vortex 2 at $\z$. Then $\rho_{1}=q\dxy$, while the scalar field due to
$\rho_{2}$ is $\psi_{2}=q\kn(\mu|\x-\z|)/2\pi$. Hence,
\beq
\label{34}
L_{\psi}=\inx\frac{q^{2}}{2\pi}\dxy\kn(\mu|\x-\z|)
=\frac{q^{2}}{2\pi}\kn(\mu|\y-\z|).
\eeq
The magnetic interaction is similar: $j_{(1)}^{0}=0$, $\jv_{(1)}=
-m\k\times\D\dxy$ while $A_{(2)}^{0}=0$, $\A_{(2)}=-m\k\times\D\kn(|\x-\z|)$,
so
\bea
L_{A}&=&\inx\frac{m^{2}}{2\pi}[\k\times\D\dxy]\cdot[\k\times\D\kn(|\x-\z|)]
\nonumber \\
&=& -\frac{m^{2}}{2\pi}\Delta_{y}\kn(|\y-\z|) \nonumber \\
\label{35}
&=& -\frac{m^{2}}{2\pi}\kn(|\y-\z|)
\eea
using (\ref{29}) with $\y\neq\z$. The total interaction Lagrangian is a 
function of $|\y-\z|$ only, so we interpret $-L_{\rm int}$ as the potential 
energy of the interaction,
\beq
\label{36}
U=\frac{1}{2\pi}[m^{2}\kn(r)-q^{2}\kn(\mu r)]
\eeq
where $r$ is the vortex separation, that is 
$\rv=r(\cos\vartheta,\sin\vartheta):=\y-\z$. This is the same potential as 
found in \cite{BetRiv}, but we arrived at it via a different route.

This potential is consistent with the partition into type I, critical and
type II regimes. The central force due to $U$ is
\beq
\label{37}
-U'(r)=\frac{1}{2\pi}[m^{2}\ko(r)-\mu q^{2}\ko(\mu r)].
\eeq
If $\mu<1$, then $\ko\rightarrow 0$ at large $r$ faster than $\ko(\mu r)$,
so scalar attraction dominates over magnetic repulsion and the force is 
negative, consistent with type I behaviour. If $\mu>1$, the reverse is true
and the force is positive at large $r$, consistent with type II behaviour.
Potentials for $\mu^{2}=0.4$ (type I) and $\mu^{2}=2.0$ (type II) are plotted 
in figure 1.
At $\mu=1$, $m\equiv q$, as explained in section \ref{sec:va} so $U\equiv 0$
and there is no net force at all.  This consistency at large $r$ emerges
regardless of the specific values of $m$ and $q$ away from $\mu=1$, and may
be attributed to the inverse relationship between a field's mass and its
range. At moderate $r$, the $\mu$ dependance of $q/m$ becomes important. Given
that $\ko$ is a strictly decreasing function, it is clear from (\ref{37}) that
there exists a unique critical point of $U$ for each $\mu\neq 1$ if and only
if $m/q>\sqrt{\mu}$ when $\mu<1$ and $m/q<\sqrt{\mu}$ when $\mu>1$. Our 
numerical work suggests that $m/q$ easily passes these criteria. The 
rightmost column of table 1 presents the approximate critical vortex separation
$r_{c}$ for each value of $\mu^{2}$. No such critical points were found by
Rebbi and Jacobs \cite{Reb}, who obtained approximate static intervortex 
potentials by numerically minimizing the potential energy functional subject to
the constraint that $\phi$ has two simple zeros separated by a given distance. 
So these equilibria are probably artifacts of our approximation, which we take
to break down for $r\leq r_{c}$. Of course, some kind of breakdown is to be 
expected: 
vortices are not point particles, as in our picture, and when they approach one
another closely enough their overlap produces significant effects.

\vbox{
\centerline{\epsfysize=3truein
\epsfbox[42 197 546 583]{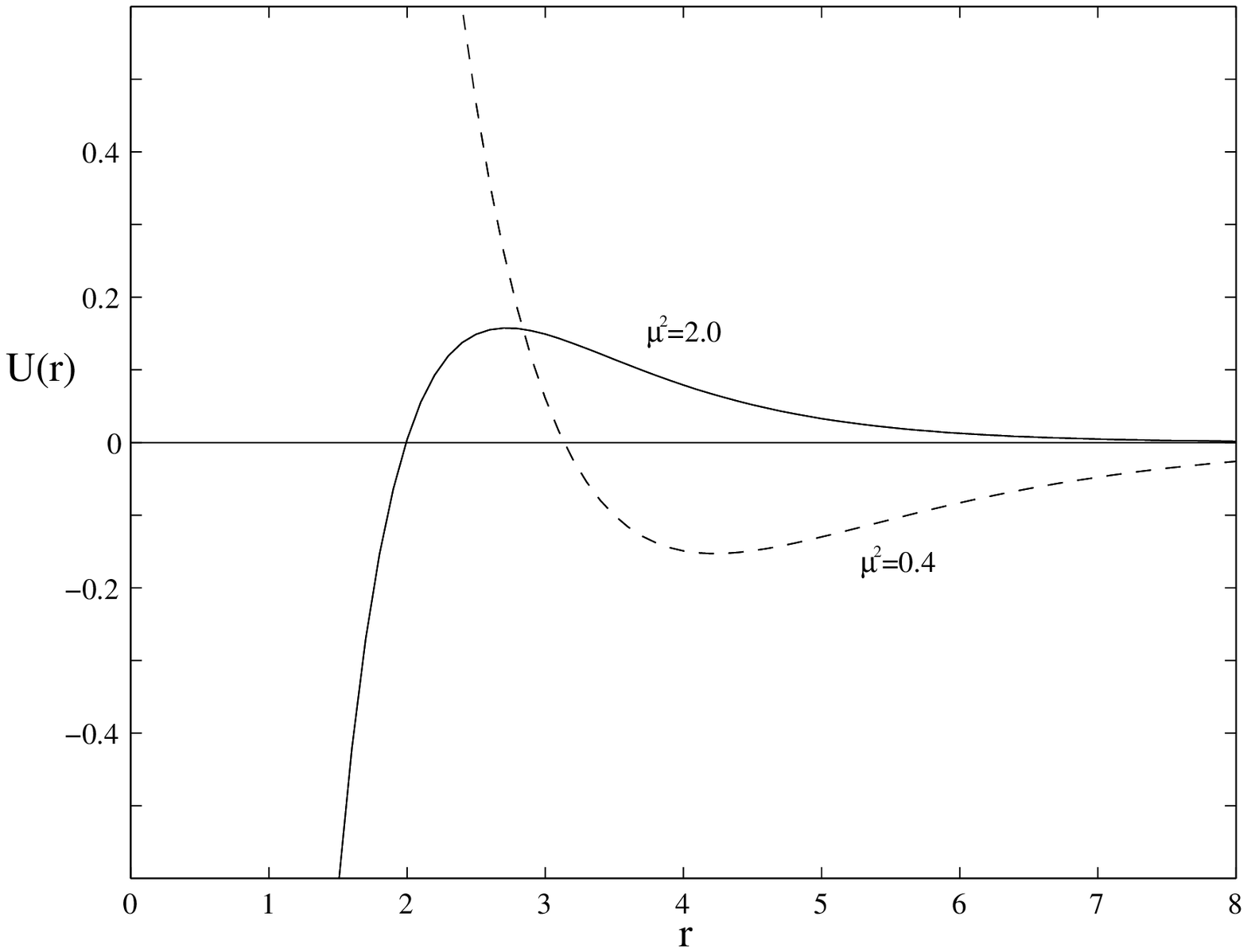}}
\centerline{\it Figure 1: The potential function ${\cal U}(r)$ for $\mu^{2}=0.4$ and 
$\mu^{2}=2.0$.}
}
\vspace{0.5cm}

\section{Type II vortex scattering}
\label{sec:tvs}

The interaction potential $U$ provides a very simple model of two-vortex
dynamics: the dynamics of two point particles each of mass $M$ (the energy of
a single vortex at rest, a $\mu^{2}$ dependent quantity) interacting via the
potential (\ref{36}). Ignoring the (trivial) centre of mass motion, the 
Lagrangian of such a mechanical system is
\beq
\label{38}
L=\frac{1}{4}M(\dot{r}^{2}+r^{2}\dot{\th}^{2})-U(r),
\eeq
since the reduced mass of the system is $M/2$. This is a manifestly bad model
if $\mu=1$, because it would predict that there is no scattering at all, in
conflict with the results of numerical simulations \cite{Mye} and the geodesic
approximation \cite{Sam}. Away from critical coupling, one might expect the
potential $U$ to dominate over velocity dependent corrections, at
least at moderately low speeds, so the above model, although simple, may give
a good quantitative account of  long range vortex interactions. 
We choose to study type~II
vortices because these provide a simple, clear-cut dynamic problem: vortex
scattering. Type I dynamics is slightly more complicated in that vortices
can scatter or form bound states depending on the initial conditions. The
coupling chosen for the type~II numerical simulations of \cite{Mye} is
$\mu^{2}=2$, a choice which we follow for purposes of comparison. In the 
Lagrangian  (\ref{38}), the constants $q$ and $m$ are already known
for $\mu^{2}=2$, but the vortex mass $M$ is not. Rather than attempt to 
calculate $M$ from our numerical solution, we use the careful numerical 
analysis of Rebbi and Jacobs \cite{Reb}. Unfortunately, they found $M$ for
each of a regular sequence of $\mu$ values, rather than $\mu^{2}$ values, so
the $\mu=\sqrt{2}$ value is not quoted. However, a graph of $\mu$ against
$M$ is very nearly linear, so we use linear interpolation to estimate the
$\mu=1.41421\ldots$ mass from the $\mu=1.4$ and $\mu=1.5$ masses given. The
result is $M=1.51230\pi$. The potential for this coupling is plotted in figure
1.

From the plot of $U(r)$ we see that all trajectories
which do not encroach on the interior region $r<r_{c}$ 
are scattering trajectories. By time-translation and rotational
symmetries, we can, without loss of generality, take the point of closest
approach (at which $r=r_{0}$ say) to lie on the
$\th=0$ ray and occur at time $t=0$. It is then straightforward to show that
$\lim_{t\rightarrow\infty}\th$ is
\beq
\label{39}
\th_{\infty}=J^{2}\int_{r_{0}}^{\infty}\frac{dr}{r^{2}}
\left[\frac{4}{M}(U(r_{0})-U(r))+J^{2}\left(\frac{1}{r_{0}^{2}}-\frac{1}{r^{2}}
\right)\right]^{-\hf}
\eeq
where $J=r^{2}\dot{\th}$ is the conserved angular momentum conjugate to $\th$.
The deflection angle $\Theta$ is $\pi-2\th_{\infty}$. Solving
the scattering problem then amounts to numerically approximating this integral.
Note that there is an integrable
singularity in the integrand at $r=r_{0}$. This presents no
problem in principle, but it
must be treated carefully in any numerical algorithm. Schematically, we handle
the integral as follows,
\bea
\th_{\infty} &=& \int_{r_{0}}^{r_{0}+\delta}+\quad\int_{r_{0}+\delta}^{\Delta}+
\quad\int_{\Delta}^{\infty} \nonumber \\
\label{40}
&\approx& \th_{\delta}\quad +\quad\th_{\rm{\scriptscriptstyle  NC}}\quad 
+\quad\th_{\Delta},
\eea
where $\delta$ is small ($\delta=0.1$) and $\Delta$ is large ($\Delta=15$).
The contribution $\th_{\delta}$ is calculated by Taylor expansion of the
integrand about $r=r_{0}$, while $\th_{\rm num}$ is evaluated using the
Newton-Cotes rule. At large $r$ the potential falls off exponentially, so for
$r>\Delta$ we set $U\equiv 0$ and calculate $\th_{\Delta}$ in the
free vortex approximation.

To make comparison with the numerical simulations described in \cite{Mye} we
calculate $\Theta$ as a function of impact parameter $b$ for scattering at 
impact speeds $v_{\infty}=0.1,0.2,0.3$ and $0.4$. The connexion between
$(b,v_{\infty})$ and $(r_{0},J)$, the parameters used in (\ref{39}) is found
using energy and angular momentum conservation.    
One might worry
that at high $v_{\infty}$ and low $b$ the vortices will 
penetrate the $r<r_{c}$ zone and become unrealistically captured. In fact even
in a head on collision, the speed required for this is greater than $0.4$, so
the problem is never encountered. The results are shown in
figure 2. As one would expect, the approximation fares reasonably well for
large impact parameters and moderate speeds, but less well in scattering 
processes where the vortices approach one another closely. The fit to the 
numerical simulations of \cite{Mye}
could be improved by adjusting the values of $q$ and $m$, a procedure which we 
eschew on the grounds that it would corrupt the deductive nature of the model.

\vbox{
\centerline{\epsfysize=3truein
\epsfbox[54 197 546 590]{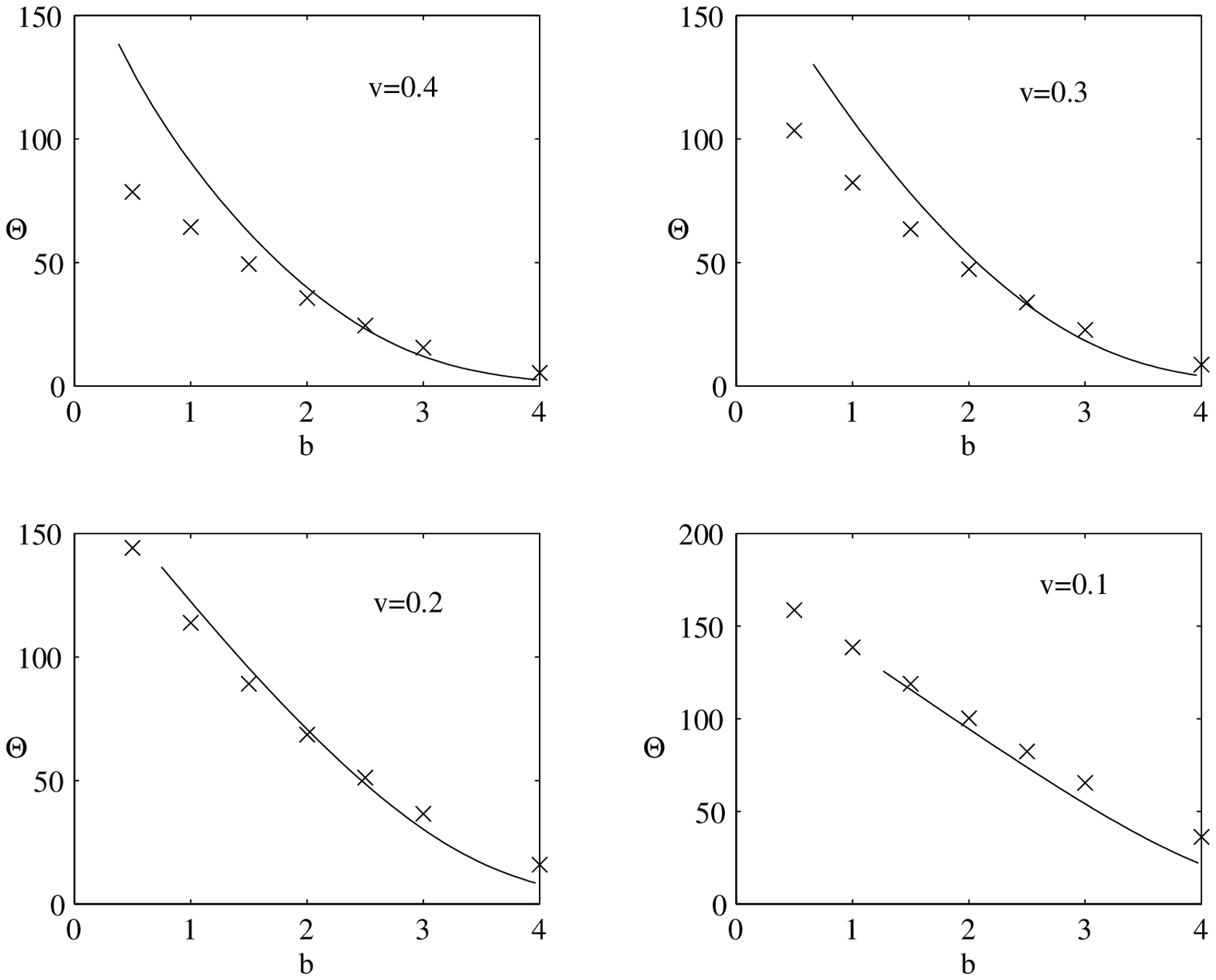}}
\noindent {\it Figure 2: The scattering of $\mu^{2}=2$ (type II) vortices: deflection
angle $\Theta$ versus impact parameter $b$ at four different impact speeds.
The solid curves were produced using the point source approximation, the
crosses by numerical simulation of the full field equations \protect\cite{Mye}.}
}
\vspace{0.5cm}

\section{Conclusion}
\label{sec:con}

In this paper we have presented a point source formalism for long range vortex
dynamics. We used this framework to rederive the static intervortex potential
from a new perspective and solved the scattering problem for $\mu^{2}=2$ 
vortices, finding reasonable agreement with numerical simulations, despite the 
simplicity of the mechanical model. It would be straightforward to apply the 
method to
other situations of interest: to derive the asymptotic forces between a
vortex-antivortex pair, or higher winding conglomerations (in the type~I
regime), or larger collections of vortices for example. 

A less straightforward extension of the present work concerns the long range
interactions of critically coupled vortices. These exert no forces on one
another when at rest, but {\em do} affect one another when in relative motion.
So the scattering of critically coupled vortices is highly nontrivial, as
evidenced by numerical simulations \cite{Mye} and a partly numerical
implementation of the geodesic approximation \cite{Sam}. In the case of
Yang-Mills-Higgs theory, another field theory the scattering of whose
critically coupled solitons (BPS monopoles) has been extensively studied
using the geodesic approximation, Manton has devised a method for finding
long range {\em velocity dependent} forces within the point particle
approximation \cite{Man4}. The idea is to calculate the interaction of one
point source with the retarded potential generated by another moving along
some trajectory. For BPS monopoles, this led to a formula for the asymptotic
metric on the two monopole moduli space, which turned out to be in precise
agreement with the asymptotic form of the Atiyah-Hitchin metric \cite{AtiHit}.
The method has been adapted to several models \cite{Sch,GibRub}, but always 
where the
linearized theory is massless. In the vortex case, the linearized theory is
massive, so the nontrivial problem is to find a suitable substitute for
ordinary retarded potentials (field disturbances no longer travel uniformly
at the speed of light, so standard retarded potentials are not appropriate).
If velocity dependent intervortex forces can be derived by this means, one
could deduce the asymptotic form of the metric on the two vortex moduli
space, at present known only numerically.

\newpage
\noindent
{\bf Acknowledgments:} I would like to thank Bernd Schroers for many valuable
conversations. This work developed partly while I was a
research student at the University of Durham, England, financially supported
by the UK Particle Physics and Astronomy Research Council.


\begin{thebibliography}{xx}

\bibitem{Hig} P.W. Higgs,
{\sl Phys. Rev.} {\bf 145} (1966) 1156.

\bibitem{Vil} A. Vilenkin,
{\sl Phys. Rep.} {\bf 121} (1985) 263.

\bibitem{BetRiv} L.M.A. Bettencourt and R.J. Rivers,
{\sl Phys. Rev.} {\bf D51} (1995) 1842.

\bibitem{NieOle} H.B. Nielsen and P. Olesen,
{\sl Nucl. Phys.} {\bf B61} (1973)  45.

\bibitem{AbrSte} M. Abramowitz and I.A. Stegun (eds),
{\sl Pocketbook of Mathematical Functions}
(Verlag Harri Deutsch, Frankfurt, Germany 1984).

\bibitem{Bog} E.B. Bogomol'nyi,
{\sl Sov. J. Nucl. Phys.} {\bf 24} (1976) 449.

\bibitem{Sam} T.M. Samols,
{\sl Commun. Math. Phys.} {\bf 145} (1992) 149.

\bibitem{Reb} L. Jacobs and C. Rebbi,
{\sl Phys. Rev.} {\bf B19} (1979) 4486.

\bibitem{Mye} E. Myers, C. Rebbi and R. Strilka,
{\sl Phys. Rev.} {\bf D45} (1992) 1355.

\bibitem{Man4} N.S. Manton,
{\sl Phys. Lett.} {\bf 154B} (1985) 397 and {\sl Phys. Lett.} {\bf 157B} (1985)
475 (errata).

\bibitem{AtiHit} M.F. Atiyah and N.J. Hitchin,
{\sl The Geometry and Dynamics of Magnetic Monopoles}
(Princeton University Press, Princeton, USA, 1988).

\bibitem{Sch} B.J. Schroers,
{\sl Z. Phys.} {\bf C61} (1994) 479.

\bibitem{GibRub} G.W. Gibbons and P.J. Ruback,
{\sl Phys. Rev. Lett.} {\bf 57} (1986) 1492.




\end{thebibliography}
\end{document}